\documentclass[aps,physrev,twocolumn,superscriptaddress,amsfonts,amssymb,amsmath]{revtex4-2}

\usepackage{graphicx}
\usepackage[most]{tcolorbox}    	
\usepackage[frozencache,cachedir=.]{minted}
\usepackage{siunitx}
\usepackage{hyperref}
\usepackage{mhchem}
\usepackage[capitalize]{cleveref}

\tcbuselibrary{minted,skins,breakable}

\newtcblisting{yamlcodebox}[2][]{%
  enhanced,
  colback=white,
  colframe=black!15,
  boxrule=0.3pt,
  arc=1mm,
  listing only,
  listing engine=minted,
  minted language=YAML,
  minted options={
    fontsize=\footnotesize,
    style=friendly,
    mathescape,
    escapeinside=||
  },
  title=#2,
  colbacktitle=black!3,
  coltitle=black,
  fonttitle=\bfseries\footnotesize,
  box align=top
  #1
}

\newtcblisting{textcodebox}[2][]{%
  enhanced,
  colback=white,
  colframe=black!15,
  boxrule=0.3pt,
  arc=1mm,
  listing only,
  listing engine=minted,
  minted language=text,
  minted options={
    fontsize=\footnotesize,
    style=friendly,
    mathescape,
    escapeinside=||
  },
  title=#2,
  colbacktitle=black!3,
  coltitle=black,
  fonttitle=\bfseries\footnotesize,
  #1
}

\newtcblisting{pythoncodebox}[1][]{%
  enhanced,
  colback=white,
  colframe=black!15,
  boxrule=0.3pt,
  arc=1mm,
  listing only,
  listing engine=minted,
  minted language=python,
  minted options={
    fontsize=\footnotesize,
    style=friendly,
    mathescape
  },
  colbacktitle=black!3,
  coltitle=black,
  fonttitle=\bfseries\footnotesize,
  #1
}

\begin{document}


\title{PACSim: A Flexible Simulation Framework for \\
Polymer-Attenuated Coulombic Self-Assembly}


\author{Philipp H\"ollmer}
\affiliation{Department of Chemistry, New York University, New York, NY 10003, USA}
\affiliation{Simons Center for Computational Physical Chemistry, New York University, New York, NY 10003, USA}

\author{Nicole Smina}
\thanks{These authors contributed equally.}
\affiliation{Department of Chemistry, New York University, New York, NY 10003, USA}

\author{John~P.~Marquardt}
\thanks{These authors contributed equally.}
\affiliation{Department of Chemistry, New York University, New York, NY 10003, USA}

\author{Michael~S.~Chen}
\affiliation{Department of Chemistry, New York University, New York, NY 10003, USA}
\affiliation{Simons Center for Computational Physical Chemistry, New York University, New York, NY 10003, USA}

\author{Steven van Kesteren}
\affiliation{Department of Chemistry, New York University, New York, NY 10003, USA}
\affiliation{Department of Information Technology and Electrical Engineering, ETH Zurich, 8092 Zurich, Switzerland}

\author{Stefano Sacanna}
\affiliation{Department of Chemistry, New York University, New York, NY 10003, USA}

\author{Glen M.~Hocky}
\email[]{hockyg@nyu.edu}
\affiliation{Department of Chemistry, New York University, New York, NY 10003, USA}
\affiliation{Simons Center for Computational Physical Chemistry, New York University, New York, NY 10003, USA}


\date{\today}

\begin{abstract}
Polymer-Attenuated Coulombic Self-Assembly (PACS) is a flexible experimental approach for generating crystals from simple colloidal building blocks. 
The central components are charged spherical particles coated with a polymer brush that prevents irreversible aggregation. 
Whether oppositely charged colloids crystallize, and which structures they form, depends on several factors, including colloid concentration, charge, and size, as well as the salt concentration of the solution. 
Molecular dynamics (MD) simulations are a powerful tool for predicting the outcomes of PACS assembly experiments and also provide particle-level insight into the assembly processes. 
Here, we present an open-source simulation framework, PACSim, that enables MD simulation studies of assembly by PACS across a range of experimentally relevant scenarios. 
PACSim is built on top of OpenMM, a flexible MD simulation framework that readily supports the implementation of different interaction potentials, as well as integration with other tools such as enhanced-sampling and machine-learning frameworks. 
We describe the motivation for PACSim, outline its features, report methodological advancements inspired by this framework, and provide examples of its use.
\end{abstract}


\maketitle

\section{Introduction}
Colloidal particles provide an experimental platform that offers deep insight into the process of crystallization, as their assembly can be observed in real time and space with simple optical microscopy~\cite{gasser2001real,Hueckel2021,Zang2024}. Moreover, crystals formed from such particles are of technological interest for optical devices because their lattices have spacings on the same length scale as visible light~\cite{Ducrot2017,Wang2017}. It has therefore been of both fundamental and practical interest to optimize routes for forming large, defect-free crystals from colloidal building blocks. In this context, computational modeling has played an important role in probing the phase space of accessible structures for such systems~\cite{Zhang2005,damasceno2012,dhakal2013growth,fang2020two,Hueckel2020,dijkstra2021predictive,Zang2025}, as the interactions between these large particles can often be described using simple, physics-informed pair potentials~\cite{hunter2001foundations,Israelachvili} which enable simulations on timescales relevant to crystallization.

An early strategy for inducing tunable interactions between colloidal particles exploited the hybridization of complementary DNA strands grafted onto particle surfaces, enabling the formation of different binary crystal structures by varying particle sizes or size ratios~\cite{Mirkin1996,Alivisatos1996,Rogers2016}. In such systems, interactions can be switched on and off sensitively through small temperature changes. 

An alternative approach mimics naturally occurring ionic crystals by combining colloids with positive and negative surface charges. This strategy takes advantage of the fact that colloidal particles often possess an innate surface charge resulting from their synthesis~\cite{leunissen2005ionic,Hueckel2020,Hueckel2021}. A central challenge, however, is that like-charge repulsion is required to prevent aggregation driven by strong dispersion interactions. In the first reported work in this area, aggregation was mitigated through the use of dielectric-matched organic solvents and custom-designed stabilizers \cite{leunissen2005ionic}. However, the reliance on specialized solvents and stabilizers, combined with the absence of a strategy to permanently fix the assembled structures, has limited the broader applicability of this approach.

To generalize the formation of ionic colloidal crystals, \citet{Hueckel2020} introduced the Polymer-Attenuated Coulombic Self-Assembly (PACS) strategy, which enables the formation of ionic crystals from a wide range of charged colloidal building blocks in water. Electrostatic interactions between the charged colloids are well described by an interaction whose strength decays exponentially over the Debye length $\lambda_D$, which can be controlled by the salt concentration of the solution. Additionally, a surface-adhering triblock copolymer forms a hydrophilic brush that prevents particles from coming into direct contact. When the brush length is comparable to $\lambda_D$, oppositely charged particles experience a short-range attractive potential whose depth can be tuned via the salt concentration. Importantly, crystals can subsequently be fixed by dialyzing out the salt, which induces strong van der Waals contacts and yields robust structures suitable for scanning electron microscopy (SEM). From a computational perspective, the pairwise PACS interaction can be modeled by combining screened electrostatics with polymer-brush repulsion, enabling molecular dynamics (MD) simulations to predict crystal structure formation as a function of particle size, surface charge, and $\lambda_D$. These simulations have reproduced experimentally observed structures~\cite{Hueckel2020,Zang2024} and revealed various classical and non-classical, as well as homogeneous and heterogeneous nucleation pathways~\cite{Zang2025,van2026light}.

Several mature MD simulation engines with distinct strengths can be used for the computational modeling of colloidal self-assembly, including HOOMD-blue~\cite{hoomd}, LAMMPS~\cite{thompson2022lammps}, and OpenMM~\cite{eastman2023openmm}. Our previous studies employed HOOMD-blue, where colloidal pair interactions were implemented via tabulated potentials~\cite{Hueckel2020,Zang2024,Zang2025,Gales2025}. HOOMD-blue is a highly efficient GPU-accelerated MD engine that supports flexible topologies, a feature we have exploited in studies of self-assembly involving explicit inducible binders~\cite{mitra2023coarse,shu2024mesoscale,holoman2026simulating}.

In this paper, we introduce PACSim, a simulation framework for PACS-based colloidal assembly built on OpenMM~\cite{eastman2023openmm}. PACSim leverages several capabilities of OpenMM that are essential for advanced crystallization studies. First, OpenMM provides native integration with the PLUMED library for enhanced sampling~\cite{plumed-consort-2019}, which offers a broad range of collective variables for crystallization analysis~\cite{tribello2025tutorials}. (It also provides straightforward coupling to PyTorch for machine-learning-based approaches~\cite{pytorch}, which we plan to leverage in future studies.) Second, its expression-based definition of custom forces enables rapid implementation of new pair and external potentials without tabulation, avoiding the need to precompute interaction tables and naturally allowing for spatial and temporal dependence. A highly optimized CUDA backend ensures efficient and scalable performance comparable to HOOMD-blue, and CPU and OpenCL backends also allow for compatibility across architectures without modification. Finally, OpenMM offers a first-class Python interface, facilitating tight integration with the broader Python ecosystem for system construction and trajectory analysis.

Other MD engines, including LAMMPS and HOOMD-blue, provide overlapping functionality with OpenMM and could, in principle, serve as drivers within PACSim. For instance, LAMMPS supports integration with both PLUMED and PyTorch, and all three engines can interface with the enhanced-sampling framework pySAGES~\cite{pysages}. Nevertheless, OpenMM currently offers a particularly convenient combination of flexible, GPU-accelerated custom force definitions, native Python control, and direct interoperability with the tools required for advanced simulations of PACS crystallization. For these reasons, it provides the foundation for the present implementation of PACSim. 

Beyond the capabilities of the underlying MD engine, PACSim provides infrastructure tailored to typical workflows in colloidal crystallization studies. It includes implementations of the PACS interactions, together with related potentials such as depletion forces, gravitational effects, and substrate interactions. PACSim also includes tools for generating initial configurations, including crystalline or gaseous states and systems containing immobile substrate particles. Non-spherical particles can be represented as rigid clusters of spheres through the use of constraints. In addition, key interaction parameters such as the Debye length $\lambda_D$ or the temperature $T$ can be varied dynamically during a simulation, optionally based on the state of the system. All simulations are specified through simple configuration files, eliminating the need for direct code modification. In this way, PACSim complements the MD engine by providing a reproducible and extensible framework for colloidal crystallization. PACSim has already been used to study systems in which colloidal surface charges respond dynamically to light in the presence of a photo acid~\cite{van2026light}.

In addition to describing the features of PACSim, we illustrate its use with a number of physical examples, many of which have not previously been reported in our earlier studies of crystallization with PACS. These include an automated dialysis protocol for finding optimal conditions for crystallization, seeded growth to assess kinetic and thermodynamic stability, competition between crystal types to compare their relative stability, and the use of enhanced-sampling simulations to study crystal stability under conditions where unbiased simulations do not crystallize on accessible timescales.

The remainder of this paper is organized as follows. In \cref{sec:theory}, we present the theoretical basis of the interactions implemented in PACSim, whose derivations are not readily available elsewhere. In \cref{sec:implementation}, we describe the structure of PACSim and its implementation with OpenMM. In \cref{sec:cookbook}, we provide concrete usage examples. Finally, \cref{sec:conclusions} summarizes our findings and discusses potential future extensions of PACSim.

\section{PACS Potentials}
\label{sec:theory}

Starting with Ref.~\cite{Hueckel2020}, we have modeled pairwise PACS interactions as the sum of two potentials that depend on the surface-to-surface distance $h$ between spherical colloids. The first contribution is a polymer-brush repulsion $V_P(h)$, which can be derived within the Alexander--de Gennes model~\cite{Alexander,DeGennes}. The second is an electrostatic interaction $V_E(h)$, analogous to the screened Coulomb potential appearing in Derjaguin--Landau--Verwey--Overbeek (DLVO) theory~\cite{Derjaguin,Verwey}. In \cref{sec:polymer,sec:electro}, we collect more detailed derivations of both potentials, which are not typically presented together in the literature.

\begin{figure*}[t]
\centering
\includegraphics{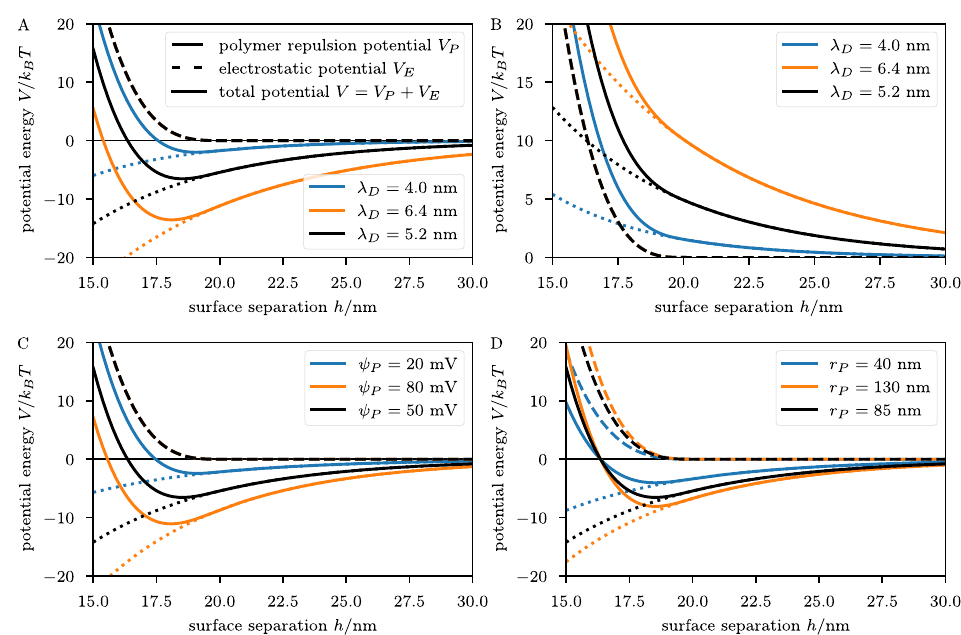}
\caption{Tuning parameters of the PACS potential $V=V_P+V_E$, consisting of the electrostatic interaction $V_E$ and the polymer-brush repulsion $V_P$. We plot the pair potential between two spherical colloids. Unless specified otherwise, parameters are fixed to typical values of the brush density $\sigma=\qty{0.09}{\per\nano\meter\squared}$, brush length $L=\qty{10}{\nano\meter}$, relative permittivity $\varepsilon=80$, temperature $T=\qty{298}{\kelvin}$, Debye length $\lambda_D=\qty{5.2}{\nano\meter}$, particle radii $r_P=\qty{85}{\nano\meter}$ and $r_N=\qty{105}{\nano\meter}$, and surface potentials $\psi_P=\qty{50}{\milli\volt}$ and $\psi_N=\qty{-50}{\milli\volt}$. (A)~Varying the Debye length $\lambda_D$ changes the position and depth of the PACS potential minimum for oppositely charged particles. (B)~For a pair of identical positively charged particles with parameters $r_P$ and $\psi_P$, the electrostatic interaction $V_E$ becomes repulsive. (C)~For oppositely charged particles, changing the surface potential $\psi_P$ shifts the position and depth of the PACS potential minimum. (D)~The particle radius $r_P$ affects the strength of the polymer-brush repulsion $V_P$ and electrostatic potential $V_E$.}
\label{fig:pacs}
\end{figure*}

The polymer-brush repulsion $V_P(h)$ prevents particles from approaching separations $h$ where van der Waals attraction (which is not modeled explicitly here) would lead to irreversible aggregation. Together with the electrostatic attraction $V_E(h)$ between oppositely charged colloids, PACS gives rise to short-range attractive ionic bonds whose strength can be tuned through parameters of the colloids and the electrolyte solution, most importantly the Debye length $\lambda_D$ (see \cref{fig:pacs}).

PACSim also implements a depletion interaction between spherical colloids with different radii that follows from the Asakura--Oosawa model~\cite{Asakura,Vrij}, which is commonly used to assemble colloidal particles~\cite{dijkstra1999phase,dijkstra2001computer,sacanna2010lock} and has been combined with PACS to enable light-driven crystallization through use of a photoacid \cite{van2026light}. The derivation of the attractive depletion potential $V_D(h)$ is presented in \cref{sec:depletion}. 

\subsection{Polymer-Brush Repulsion}
\label{sec:polymer}

In the PACS system, polymer brushes grafted onto the surfaces of the colloids act as steric spacers whose repulsive interactions prevent particles from entering the attractive van der Waals regime. The functional form of this repulsion potential can be derived within the Alexander--de Gennes model~\cite{Alexander,DeGennes}. Here, we follow the derivation presented by \citet{Likos2000}.

The Alexander--de Gennes expression for the repulsive pressure $P_P$ between two planar surfaces coated with uniform polymer brushes of thickness $L$, separated by an interplate distance $D$, is given by \citet{DeGennes}:
\begin{equation}
\label{eq:pressure}
    P_P(D)=\frac{k_B T}{s^3} \left[ \left(\frac{2L}{D}\right)^{\frac{9}{4}} - \left(\frac{D}{2L}\right)^{\frac{3}{4}} \right]\text{ for $D<2L$}.
\end{equation}
For $D \geq 2L$, the pressure vanishes. In \cref{eq:pressure}, $s$ denotes the average distance between grafting points on the surfaces. The first osmotic term arises from the increase in polymer concentration as $D$ decreases, while the second term reflects the change in elastic energy under compression~\cite{OShea}.
The pressure $P_P(D)$ is related to the free energy per unit area $W_P(D)$ via $P_P(D)=-\partial W_P(D)/\partial D$. Imposing the boundary condition $W_P(D)=0$ for $D\geq2L$, we obtain
\begin{equation}
\begin{split}
W_P(D)&=\int_D^{2L}P_P(x)\,\mathrm{d}x \\
&= \frac{8k_BTL}{35s^3}\left[7\left(\frac{2L}{D}\right)^{\frac{5}{4}} + 5 \left(\frac{D}{2L}\right)^\frac{7}{4} - 12\right].
\end{split}
\end{equation}

To obtain the effective pair potential between two spherical colloids with radii $r_i$ and $r_j$, each coated with polymer brushes of length $L$ and separated by a surface-to-surface distance $h$, we employ the Derjaguin approximation~\cite[Section~11.5]{Israelachvili}. Within this approximation, the force $F_P(h)$ between the two bodies is given by $F_P(h)=\pi R_{ij}W_P(h)$, where $R_{ij}=2/(r_i^{-1}+r_j^{-1})$ is the effective radius. The corresponding interaction potential then follows for $0\leq h\leq 2L$ as
\begin{equation}
\label{eq:brush}
\begin{split}
V_P(h) &= \pi R_{ij}\int_h^{2L} W_P(x)\,\mathrm{d}x \\
&=\frac{16\pi k_BT R_{ij}L^2}{35s^3}\Bigg\{ 
28\left[ \left(\frac{2L}{h}\right)^{\frac{1}{4}} - 1 \right] \\
&\hphantom{=\frac{16\pi k_BT R_{ij}L^2}{35s^3}\Bigg\{}+\frac{20}{11}\left[ 1-\left(\frac{h}{2L}\right)^\frac{11}{4}\right] \\
&\hphantom{=\frac{16\pi k_BT R_{ij}L^2}{35s^3}\Bigg\{}+ 12 \left[ \frac{h}{2L}-1\right]
\Bigg\},
\end{split}
\end{equation}
where $h>0$, and we imposed $V_P(h) = 0$ for $h\geq 2L$. With the introduction of the brush density $\sigma=1/s^2$, \cref{eq:brush} recovers the polymer-brush potential used in \citet{Hueckel2020}.

The Derjaguin approximation assumes that the surface-to-surface distance between the two colloidal spheres is small compared to their radii of curvature, i.e., $r_i\gg h$ and $r_j\gg h$~\cite[Section~11.5]{Israelachvili}. This condition is naturally satisfied in the PACS systems we are interested in, where the brush length $L$---and thus the relevant range of $h$---is typically on the order of $\qty{10}{\nano\meter}$ while the radii are on the order of $\qty{100}{\nano\meter}$.

\subsection{Electrostatic Interaction}
\label{sec:electro}

Near a charged surface in an electrolyte solution, there is an accumulation of counterions (ions of opposite charge to the surface) and a depletion of co-ions, forming a ``diffuse electric double layer''~\cite{{Israelachvili}}.
In the limit of low surface potentials, the diffuse layer behaves like a parallel-plate capacitor with an effective plate separation given by the Debye length $\lambda_D$, which is set by the electrolyte properties~\cite[Section~14.14]{Israelachvili}.

More generally, the interaction free energy per unit area $W_E(D)$ between two planar surfaces separated by a distance $D$, bearing low constant surface potentials $\psi_i$ and $\psi_j$, in a 1:1 electrolyte solution can be written as~\cite[Section~14.18]{Israelachvili}:
\begin{equation}
\label{eq:freeel}
W_E(D)=\frac{\varepsilon_0\varepsilon\left[2\psi_i\psi_j-(\psi_i^2+\psi_j^2)\exp(-D/\lambda_D)\right]}{\lambda_D\left[\exp(D/\lambda_D)-\exp(-D/\lambda_D)\right]}.
\end{equation}
Here, all quantities are given in SI units, $\varepsilon_0$ is the vacuum permittivity, and $\varepsilon$ is the relative permittivity of the suspending medium.

The effective pair potential $V_E(h)$ between double layers on two spherical particles with effective radius $R_{ij}$ separated by a surface-to-surface distance $h$ is derived from \cref{eq:freeel} within the Derjaguin approximation by \citet{Hogg}:
\begin{equation}
\begin{split}
V_E(h) = \frac{\pi\varepsilon_0\varepsilon R_{ij}}{2}\bigg[&2\psi_i\psi_j\ln\left(\frac{1+\mathrm{e}^{-h/\lambda_D}}{1-\mathrm{e}^{-h/\lambda_D}}\right) \\
&+ \left(\psi_i^2+\psi_j^2\right)\ln\Big(1-\mathrm{e}^{-2h/\lambda_D}\Big)\bigg].
\label{eq:elect_full}
\end{split}
\end{equation}
For $h/\lambda_D\gg1$, which is enforced by the polymer-brush repulsion, this can be simplified to
\begin{equation}
\label{eq:el}
    V_E(h)\approx2\pi\varepsilon_0\varepsilon R_{ij}\psi_i\psi_j\mathrm{e}^{-h/\lambda_D},
\end{equation}
which recovers the electrostatic potential in \citet{Hueckel2020}.
As in \cref{sec:polymer}, the Derjaguin approximation is valid when the interaction acts over a length scale that is small compared to the particle radii of curvature. For the electrostatic double-layer interactions, this length scale is set by the Debye length $\lambda_D$, and we require $r_i\gg\lambda_D$ and $r_j\gg\lambda_D$. This condition is naturally satisfied in PACS systems that we study, where the spherical colloid radii are on the order of $\qty{100}{\nano\meter}$ and $\lambda_D$ is on the order of $\qty{5}{\nano\meter}$. Moreover, the expression in \cref{eq:freeel} relies on a low-potential approximation and \citet{Hogg} conclude that it is accurate for $\psi_i$ and $\psi_j$ less than $\qtyrange{50}{60}{\milli\volt}$, which is the typical range of surface potentials in PACS systems. 

The constant-potential boundary condition implicit in \cref{eq:freeel} is most questionable when the diffuse layers strongly overlap, i.e., for separations $h\lesssim\lambda_D$~\cite[Section~14.17]{Israelachvili}. In PACS, this near-contact regime is strongly suppressed by the polymer-brush repulsion which is steeply repulsive for separations below $2 L$, with the polymer brush length $L\approx\qty{10}{\nano\meter}$ (see \cref{fig:pacs}). This also justifies the large-separation expansion used in \cref{eq:el}.

\subsection{Depletion Interaction}
\label{sec:depletion}

The Asakura--Oosawa model considers colloidal particles immersed in a dilute solution of non-adsorbing depletants of radius $r_D$~\cite{Asakura,Vrij}. Because of the finite depletant size, the centers of mass of the depletants are excluded from a layer of thickness $r_D$ surrounding each colloid. When two colloids approach such that these exclusion layers overlap, the overlap volume $\mathcal{V}$ becomes accessible to
depletants, which lowers the free energy by an amount equal to the osmotic pressure $\Pi$ of the depletants times the additional free volume $\mathcal{V}$ relative to when the colloids are at large separation. The effective (attractive) depletion interaction between two colloidal particles at surface-to-surface distance $h$ can therefore be written as~\cite{Lekkerkerker}
\begin{equation}
\label{eq:ao_start}
V_D(h)=-\Pi\,\mathcal{V}(h).
\end{equation}
\Cref{eq:ao_start} holds quite generally for colloidal particles of various shapes~\cite{Petukhov}.

For an isolated spherical colloid of radius $r_i$ coated with polymer brushes of length $L$, the centers of mass of depletants are excluded from a spherical region of radius $a_i=r_i+L+r_D$.
The overlap volume of two spheres of radii $a_i$ and $a_j$ at distance $d$ is given by
\begin{equation}
\label{eq:overlap}
    \mathcal{V}(d) = \frac{\pi(a_i + a_j - d)^2[d^2 + 2d(a_i+a_j)-3(a_i - a_j)^2]}{12\,d}.
\end{equation}
Naturally, $\mathcal{V}(d)=0$ for $d\geq a_i+a_j$. \Cref{eq:overlap} can be rewritten into a function of the surface-to-surface distance $h$ by using $d=h+r_i+r_j$.

For ideal non-adsorbing depletants, the osmotic pressure is given by $\Pi=n_Dk_BT$, where $n_D$ is the depletant number density. For conditions where we can consider depletants to be hard spheres of radius $r_D$, their volume fraction is given by $\phi=n_D(4\pi r_D^3/3)$, and we can parameterize $\Pi$ in terms of $\phi$ as $\Pi = 3\phi k_BT/(4\pi r_D^3)$.

\section{PACSim}
\label{sec:implementation}

PACSim provides the \texttt{pacsim-create} utility that generates the initial colloidal configuration (see \cref{sec:pacsim-create}). This configuration then serves as input to the \texttt{pacsim-run} utility, which performs an MD simulation using OpenMM (see \cref{sec:pacsim-run}). Together, these two tools provide a flexible workflow for constructing and simulating PACS systems.
PACSim, as well as the code and data for this article, are available from \url{https://github.com/hocky-research-group/PACSim}.

\subsection{pacsim-create}
\label{sec:pacsim-create}

\begin{figure}[tb]
\centering
\includegraphics{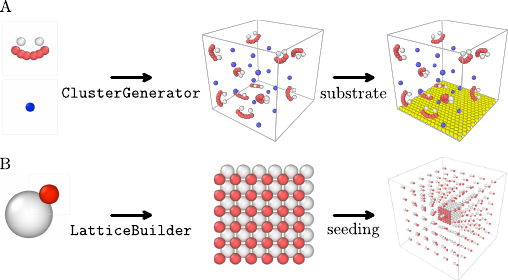}
\caption{(A)~The \texttt{ClusterGenerator} in \texttt{pacsim-create} replicates cluster templates specified in LAMMPS data files to generate an initial base configuration, after which a substrate modifier adds an explicit substrate at the bottom of the simulation box. (B)~The \texttt{LatticeBuilder} in \texttt{pacsim-create} rescales a crystal structure from a CIF file to the colloidal length scale and expands it into a supercell, which can then be used as a crystalline seed by a seeding modifier.}
\label{fig:pacsim_create}
\end{figure}

The \texttt{pacsim-create} utility generates initial particle configurations and bond topologies. It constructs systems in two stages. First, a base configuration is generated, and then second, any selected modifiers are applied to adjust the starting configuration (see \cref{fig:pacsim_create}).

All generator settings, particle properties, and modifiers are defined within a single YAML configuration file. Resulting configurations are written in the GSD format. Originally developed as the native file format for HOOMD-blue, GSD is a compact and flexible binary format that stores particle positions, velocities, bond topology, and per-particle properties such as masses, radii, and surface potentials in one file, while allowing efficient random access and straightforward interoperability with Python-based analysis tools, as well as visualization software such as Ovito~\cite{ovito} and VMD~\cite{vmd}. In PACSim, the GSD format is interfaced with OpenMM to initialize particle configurations and to store the full simulation trajectory, as well as possibly dynamic system properties, e.g., energy, temperature, or Debye length.

\begin{figure}[htb]
\begin{yamlcodebox}{configuration\_cluster\_generator.yaml}
configuration_generator: "ClusterGenerator"
configuration_generator_parameters:
  cluster_specifications: ["cluster.lmp"]
  cluster_relative_weights: [|1.0|]
  # Repeat cluster 15 times along every direction.
  lattice_repeats: [|15|, |15|, |15|]
  # Scaling for the lattice vectors in cluster.lmp.
  cluster_padding_factor: 1.0
  # Scaling for the lattice vectors after replication.
  padding_factor: 1.066666666666667
  # Randomly rotate every cluster replica.
  random_rotation: true
masses:
  # Quantities specify values with specific units.
  # They are converted to PACSim's unit system.
  # Atom types "1" and "2" appear in "cluster.lmp".
  "1": !Quantity
    unit: "dalton"
    value: 1.24898
  "2": !Quantity
    unit: "dalton"
    value: 1.0
radii:
  "1": !Quantity
    unit: "nanometer"
    value: 105.0
  "2": !Quantity
    unit: "nanometer"
    value: 97.5
surface_potentials:
  "1": !Quantity
    unit: "millivolt"
    value: -50.0
  "2": !Quantity
    unit: "millivolt"
    value: 50.0 
# Can be used to generate substrate.
initial_modifiers: null
initial_modifiers_parameters: null
# Can be used to overlay seed.
final_modifiers: null
final_modifiers_parameters: null
\end{yamlcodebox}
\caption{Configuration file for \texttt{pacsim-create} that sets up the \texttt{ClusterGenerator} to construct a base configuration by replicating a cluster template specified in the LAMMPS data file \texttt{cluster.lmp} (see \cref{fig:lammps}). The generated particle positions, masses, radii, and surface potentials are stored together with the bond topology in a GSD file.}
\label{fig:configuration_one}
\end{figure}
\begin{figure}[tb]
\begin{textcodebox}
{cluster.lmp}
|\textcolor[HTML]{60A0B0}{\textit{(Smiley cluster of Fig. 2a)}}|

8 |\textcolor[HTML]{4070A0}{atoms}|
15 |\textcolor[HTML]{4070A0}{bonds}|
2 |\textcolor[HTML]{4070A0}{atom types}|
1 |\textcolor[HTML]{4070A0}{bond types}|

|\textcolor[HTML]{60A0B0}{\textit{# Cell information.}}|
0.0    3000  |\textcolor[HTML]{4070A0}{xlo xhi}|  
0.0    3000  |\textcolor[HTML]{4070A0}{ylo yhi}|
0.0    3000  |\textcolor[HTML]{4070A0}{zlo zhi}|
0.0 0.0 0.0  |\textcolor[HTML]{4070A0}{xy xz yz}|

|\textcolor[HTML]{062873}{Atoms}|
|\textcolor[HTML]{60A0B0}{\textit{# Atom index, molecule ID, atom type, charge, x, y, z.}}|
1   0   1   0.0  0   500   500
2   0   1   0.0  0   500   -500
...

|\textcolor[HTML]{062873}{Bonds}|
|\textcolor[HTML]{60A0B0}{\textit{# Bond index, bond ID, atom index one, atom index two.}}|
1   0   1    2
2   0   1    3
...
\end{textcodebox}
\caption{Specification of a cluster template in a LAMMPS data file that gets used by the \texttt{ClusterGenerator} in \texttt{pacsim-create} to construct a base configuration (see \cref{fig:configuration_one}). The cluster is specified by cell information, atom information, and bond information. (The molecule ID, charge, and bond ID are ignored.)}
\label{fig:lammps}
\end{figure}

Two types of base configuration generators are currently implemented. Both allow large initial configurations to be constructed from structural templates containing only a small number of particles. The \texttt{ClusterGenerator} constructs systems from cluster templates specified in LAMMPS data files (see \cref{fig:configuration_one,fig:lammps} for an example of the configuration file and LAMMPS data file, as well as \cref{fig:pacsim_create}A for the generated configuration). These LAMMPS templates provide three essential pieces of information: particle positions, cell vectors (including non-orthogonal cells), and bond topology. During configuration generation, clusters are replicated along the lattice vectors of the template cell, with independently configurable repeat counts in each direction. Multiple cluster types may be specified, with weighted random selection controlling their relative abundance, and clusters can optionally be randomly rotated upon placement. Bonds defined in the template are interpreted as rigid distance constraints, allowing composite particles---such as dumbbells, tetrahedra, or other non-spherical colloids—to be represented as clusters of spheres. Padding parameters control both the minimum spacing between clusters and the distance between the outermost particles and the simulation box boundaries.

The \texttt{LatticeBuilder} (created for Ref.~\cite{vankesteren-surfactant-2026}) builds configurations from crystal structures provided in the Crystallographic Information File (CIF) format using utilities from the atomic simulation environment (ASE) Python package~\cite{hjorth2017ase} (see \cref{fig:configuration_two,fig:cif} for an example of the configuration file and CIF file, as well as \cref{fig:pacsim_create}B for the generated configuration). CIF files typically specify atomic crystal structures. Here, the atomic positions are interpreted as the positions of spherical colloidal particles (without explicit bonds). These structures are expanded into supercells with configurable repeat counts along the lattice vectors. The lattice is then uniformly scaled to eliminate particle overlaps while accounting for configurable particle radii $r_i$ and brush thickness $L$. Optionally, the scale factor can be further refined by minimizing the sum of the polymer-brush repulsion $V_P$ and electrostatic interaction $V_E$. This generator is particularly useful for MD simulations that test the stability of candidate crystal structures.

\begin{figure}[tb]
\begin{yamlcodebox}{configuration\_lattice\_builder.yaml}
configuration_generator: "LatticeBuilder"
configuration_generator_parameters:
  lattice_specification: "CsCl.cif"
  # Repeat structure 6 times along every direction.
  lattice_repeats: 6
  # Extra padding when checking for overlaps.
  radii_padding: !Quantity
    unit: "nanometer"
    value: 20.0 
  # Extra padding for the lattice vectors.
  lattice_padding: !Quantity
    unit: "nanometer"
    value: 412.61
  # Find scaling that minimizes PACS potentials.
  optimize_energy: true
  # To optimize energy, one needs access to $\lambda_D$ etc.
  run_parameters_file: "run.yaml"
masses:
  # Quantities specify values with specific units.
  # They are converted to PACSim's unit system.
  # Atom types "Cs" and "Cl" appear in "CsCl.cif".
  "Cs": !Quantity
    unit: "dalton"
    value: 1.0
  "Cl": !Quantity
    unit: "dalton"
    value: 1.0
radii:
  "Cs": !Quantity
    unit: "nanometer"
    value: 85.0
  "Cl": !Quantity
    unit: "nanometer"
    value: 105.0
surface_potentials:
  "Cs": !Quantity
    unit: "millivolt"
    value: 50.0
  "Cl": !Quantity
    unit: "millivolt"
    value: -50.0
# Can be used to generate substrate.
initial_modifiers: null
initial_modifiers_parameters: null
# Can be used to overlay seed.
final_modifiers: null
final_modifiers_parameters: null
\end{yamlcodebox}
\caption{Configuration file for \texttt{pacsim-create} that sets up the \texttt{LatticeBuilder} to construct a base configuration by replicating a crystal structure specified in the CIF file \texttt{CsCl.cif} (see \cref{fig:cif}). The generated particle positions, masses, radii, and surface potentials are stored together with the bond topology in a GSD file.}
\label{fig:configuration_two}
\end{figure}
\begin{figure}[tb]
\begin{textcodebox}{CsCl.cif}
|\textcolor[HTML]{60A0B0}{\textit{# generated using pymatgen}}|
|\textcolor[HTML]{062873}{data\_CsCl}|
|\textcolor[HTML]{062873}{\_symmetry\_space\_group\_name\_H-M}|   |\textcolor[HTML]{4070A0}{'P 1'}|
|\textcolor[HTML]{062873}{\_cell\_length\_a}|   4.14369834
|\textcolor[HTML]{062873}{\_cell\_length\_b}|   4.14369834
|\textcolor[HTML]{062873}{\_cell\_length\_c}|   4.14369834
|\textcolor[HTML]{062873}{\_cell\_angle\_alpha}|   90.00000000
|\textcolor[HTML]{062873}{\_cell\_angle\_beta}|   90.00000000
|\textcolor[HTML]{062873}{\_cell\_angle\_gamma}|   90.00000000
|\textcolor[HTML]{062873}{\_symmetry\_Int\_Tables\_number}|   1
|\textcolor[HTML]{062873}{\_chemical\_formula\_structural}|   |\textcolor[HTML]{4070A0}{CsCl}|
|\textcolor[HTML]{062873}{\_chemical\_formula\_sum}|   |\textcolor[HTML]{4070A0}{'Cs1 Cl1'}|
|\textcolor[HTML]{062873}{\_cell\_volume}|   71.14827813
|\textcolor[HTML]{062873}{\_cell\_formula\_units\_Z}|   1
|\textcolor[HTML]{062873}{loop\_}|
 |\textcolor[HTML]{062873}{\_symmetry\_equiv\_pos\_site\_id}|
 |\textcolor[HTML]{062873}{\_symmetry\_equiv\_pos\_as\_xyz}|
  1  |\textcolor[HTML]{4070A0}{'x, y, z'}|
|\textcolor[HTML]{062873}{loop\_}|
 |\textcolor[HTML]{062873}{\_atom\_type\_symbol}|
 |\textcolor[HTML]{062873}{\_atom\_type\_oxidation\_number}|
  |\textcolor[HTML]{4070A0}{Cs+}|  1.0
  |\textcolor[HTML]{4070A0}{Cl-}|  -1.0
|\textcolor[HTML]{062873}{loop\_}|
 |\textcolor[HTML]{062873}{\_atom\_site\_type\_symbol}|
 |\textcolor[HTML]{062873}{\_atom\_site\_label}|
 |\textcolor[HTML]{062873}{\_atom\_site\_symmetry\_multiplicity}|
 |\textcolor[HTML]{062873}{\_atom\_site\_fract\_x}|
 |\textcolor[HTML]{062873}{\_atom\_site\_fract\_y}|
 |\textcolor[HTML]{062873}{\_atom\_site\_fract\_z}|
 |\textcolor[HTML]{062873}{\_atom\_site\_occupancy}|
  |\textcolor[HTML]{4070A0}{Cs+}|  |\textcolor[HTML]{4070A0}{Cs0}|  1  0.00000000  0.00000000  0.00000000  1
  |\textcolor[HTML]{4070A0}{Cl-}|  |\textcolor[HTML]{4070A0}{Cl1}|  1  0.50000000  0.50000000  0.50000000  1
\end{textcodebox}
\caption{Specification of a crystal structure in a CIF file that gets used by the \texttt{LatticeBuilder} in \texttt{pacsim-create} to construct a base configuration (see \cref{fig:configuration_two}). This CIF file specifies a \ce{CsCl} crystal structure and is available in the Materials Project (identifier \texttt{mp-22865})~\cite{mp}.}
\label{fig:cif}
\end{figure}

After the base configuration has been generated, optional modifiers can be applied. A substrate modifier adds a hexagonal close-packed layer of immobile particles at the bottom of the simulation box to model an explicit substrate surface (see \cref{fig:pacsim_create}A). A seeding modifier overlays particles from an external GSD configuration, removing overlapping base particles before merging the two particle sets (see \cref{fig:pacsim_create}B). These features facilitate simulations of heterogeneous nucleation on substrates and seeded crystal growth.

\subsection{pacsim-run}
\label{sec:pacsim-run}

The \texttt{pacsim-run} utility performs MD simulations of PACS systems using OpenMM. It reads the initial particle positions, bond topology, particle properties, and simulation cell information from the GSD configuration file generated by \texttt{pacsim-create}. These data define the system state from which the simulation is initialized. In its current implementation, PACSim performs MD simulations in the canonical ($NVT$) ensemble. Through OpenMM, PACSim supports several integrators suitable for $NVT$ simulations, including Langevin and Brownian dynamics as well as Nosé--Hoover chain dynamics~\cite{Zhang2019}. Bonds that were specified in the initial configuration are interpreted as pairwise distance constraints, enabling clusters of particles to move as rigid composite objects. Distance constraints are handled by OpenMM using the Constant Constraint Matrix Approximation algorithm~\cite{Eastman2010}. This method is parallelized on GPUs and can therefore efficiently enforce large numbers of constraints. Care must be taken, however, not to apply redundant constraints that exceed the internal degrees of freedom of a particle cluster.

All simulation-specific parameters are specified in a YAML configuration file (see \cref{fig:run.yaml} for an example). This file collects general simulation settings such as the integration scheme, time step, and total number of time steps, as well as the parameters defining the interaction potentials. In PACSim, the electrostatic interaction $V_E$ and the polymer-brush repulsion $V_P$ are always present. Additional forces and their parameters can optionally be added through the same configuration file. These include, for example, gravitational forces, depletion interactions, and enhanced-sampling forces defined through the PLUMED library.

\begin{figure*}[tb]
\begin{minipage}[t]{0.49\textwidth}
\begin{yamlcodebox}{run.yaml}
# Parameters of integrator.
integrator: LangevinIntegrator
integrator_parameters:
  # Random number seed for integrator.
  randomNumberSeed: 1
  # Quantities specify values with specific units.
  # They are converted to PACSim's unit system.
  frictionCoeff: !Quantity
    unit: /picosecond
    value: 0.0016
  stepSize: !Quantity
    unit: picosecond
    value: 0.003
  # Temperature of $NVT$ integrator.
  # Copy over another entry from same YAML file.
  temperature: !Copy
    key: potential_temperature

# Parameters of PACS potentials.
# Temperature $T$ in Eq. (3).
potential_temperature: !Quantity
  unit: kelvin
  value: 298.0
# Brush density $\sigma$ in Eq. (3).
brush_density: !Quantity
  unit: /(nanometer**2)
  value: 0.09
# Brush length $L$ in Eq. (3).
brush_length: !Quantity
  unit: nanometer
  value: 10.0
# Relative permittivity $\varepsilon$ in Eq. (6).
dielectric_constant: 80.0
# Debye length $\lambda_D$ in Eq. (6).
debye_length: !Quantity
  unit: nanometer
  value: 5
# Cutoff electrostatics after $2r_\mathrm{max}+21\lambda_D$.
# Here, $r_\mathrm{max}$ is the maximum radius in the simulation.
cutoff_factor: 21.0
\end{yamlcodebox}
\end{minipage}\hfill
\begin{minipage}[t]{0.49\textwidth}
\begin{yamlcodebox}{run.yaml}
# Parameters of box walls.
wall_directions: [|true|, |true|, |true|]
# Parameter $\epsilon$ in Eq. (9).
epsilon: !Quantity
  unit: kilojoule/mole
  value: 2.477709860209665
# Parameter $\alpha$ in Eq. (9).
alpha: 1.0

# Switches for additional potentials.
use_implicit_substrate: false
substrate_wall_charge: null
use_depletion: false
use_gravity: false
use_plumed: false
plumed_script: null
# Can be used to modify parameters during run.
update_reporter: null
update_reporter_parameters: null

# Total number of time steps.
run_steps: 500000000
# Initial configuration created by pacsim-create.
initial_configuration: initial_configuration.gsd
# Time steps between storing energies and temperature.
state_data_interval: 10000
state_data_filename: state_data.csv
# Time steps between storing trajectory snapshots.
trajectory_interval: 1000000
trajectory_filename: trajectory.gsd
# Run simulation on GPU.
platform_name: CUDA
# Random number seed for initial velocities.
velocity_seed: 1
\end{yamlcodebox}
\end{minipage}
\caption{Configuration file for \texttt{pacsim-run} that runs an MD simulation in the $NVT$ ensemble of a PACS system in a confined box starting from a configuration created by \texttt{pacsim-create}. }
\label{fig:run.yaml}
\end{figure*}

The polymer-brush, electrostatic, and depletion interactions in PACSim are implemented using OpenMM's custom force framework, which allows potentials to be specified directly as analytical expressions in the form of text strings~\cite{eastman2012}. These expressions are automatically compiled by OpenMM into efficient GPU code, enabling rapid implementation and modification of interaction potentials without requiring low-level GPU programming (see \cref{fig:code}). Moreover, custom forces may depend on the actual spatial positions of the particles in addition to pairwise distances, which allows for the implementation of gravity and more complex spatially varying interactions such as those introduced experimentally through dialysis or local light illumination on a photoacid~\cite{Zang2025,van2026light}. Forces defined through PLUMED are readily integrated through an OpenMM plugin that only requires a PLUMED control script.

\begin{figure}[tb]
\begin{pythoncodebox}
from openmm import CustomNonbondedForce

# Omitting constant prefactor $2\pi\varepsilon_0\varepsilon$ for brevity.
ve = CustomNonbondedForce("""
ra * psi1 * psi2 * exp(-h / debye_length);
ra = 2 / (1 / radius1 + 1 / radius2);
h = r - radius1 - radius2
""")
ve.addGlobalParameter("debye_length", 5.0)
ve.addPerParticleParameter("radius")
ve.addPerParticleParameter("psi")    
\end{pythoncodebox}
\caption{Example Python implementation of the electrostatic interaction $V_E(h)$ from \cref{eq:el} between two spherical charged colloids using OpenMM's custom force framework. The analytical expression depends on the center-to-center distance $r$ between particles and is automatically compiled into efficient GPU code. OpenMM allows the definition of global parameters (Debye length $\lambda_D=\qty{5}{\nano\meter}$), per-particle parameters (radii $r_i$ and surface potentials $\psi_i$), and auxiliary variables within the same expression (e.g., surface-to-surface distance $h=r-r_i-r_j$).}
\label{fig:code}
\end{figure}

In OpenMM, ``reporters'' are invoked periodically at configurable time intervals during a simulation and are provided with information about the current system state. In PACSim, we implement a reporter that stores the trajectory (including positions, velocities, bonds, simulation cell information, and per-particle parameters) in a GSD file, which can be conveniently visualized using, e.g., OVITO~\cite{ovito} or VMD~\cite{vmd}. PACSim also exploits OpenMM's reporting infrastructure to optionally modify simulation parameters dynamically during an MD run in ``update reporters.'' This includes, for example, varying global parameters such as the Debye length $\lambda_D$ or integrator parameters such as the temperature $T$. Importantly, these modifications can also depend on the current time and the state of the system. For instance, $\lambda_D$ can oscillate throughout the MD simulation, or it can be increased gradually until a cluster of a target size has formed and then kept constant.

PACSim may also include substrates at the bottom of the simulation box, which can be modeled either explicitly or implicitly. In the explicit representation, \texttt{pacsim-create} generates a close-packed layer of immobile colloidal particles that interact with the mobile particles through the polymer-brush and electrostatic potentials. Alternatively, PACSim supports an implicit substrate representation, which corresponds to the interaction of the mobile particles with an effective substrate particle of infinite radius. Both approaches promote heterogeneous nucleation at the substrate surface. In addition, gravitational forces, implemented through a custom force depending on the $z$ position of the particles, can optionally be included to bias particle motion towards the substrate.

To confine particles within the simulation box, PACSim typically employs shifted Lennard--Jones walls, which are also implemented using OpenMM's custom force framework. The algebraic form is taken from HOOMD-blue~\cite{hoomd}. Specifically, with $d\geq0$ as the distance of a particle with radius $r_i$ to the wall, the shifted Lennard--Jones potential is given by
\begin{equation}
\begin{split}
V_\mathrm{SLJ}(d) =&  4\epsilon \left[\left(\frac{r_i}{d-\delta_i}\right)^{12} -
\alpha \left({\frac{r_i}{d-\delta_i}}\right)^6\right] \\ &-4\epsilon \left[\left(\frac{r_i}{r_i^c}\right)^{12} - \alpha \left(\frac{r_i}{r_i^c}\right)^6 \right].
\end{split}
\end{equation}
Here, $\epsilon$ is the well-depth parameter, $\alpha\in[0,1]$ determines the strength of the attractive part, $r_i^c=2^{1/6}r_i$, and $\delta_i=r_i-\qty{1}{\nano\meter}$. For $\alpha=1$, the potential $V_\mathrm{SLJ}(d)$ and the corresponding force go smoothly to zero at $d=r_i^c+\delta_i$. Beyond this distance, $V_\mathrm{SLJ}(d)=0$. To study bulk systems without confining walls, PACSim can alternatively apply periodic boundary conditions along an arbitrary set of directions.

\section{PACSim Cookbook}
\label{sec:cookbook}
We have created a number of examples to demonstrate the use of PACSim. All inputs and outputs for this section are available in the cookbook directory of the PACSim repository, and a single Python Jupyter notebook, \texttt{Tutorial.ipynb}, can be used to run all examples and generate outputs.

\subsection{Nucleation}

\begin{figure}
\centering
\includegraphics[width=\linewidth]{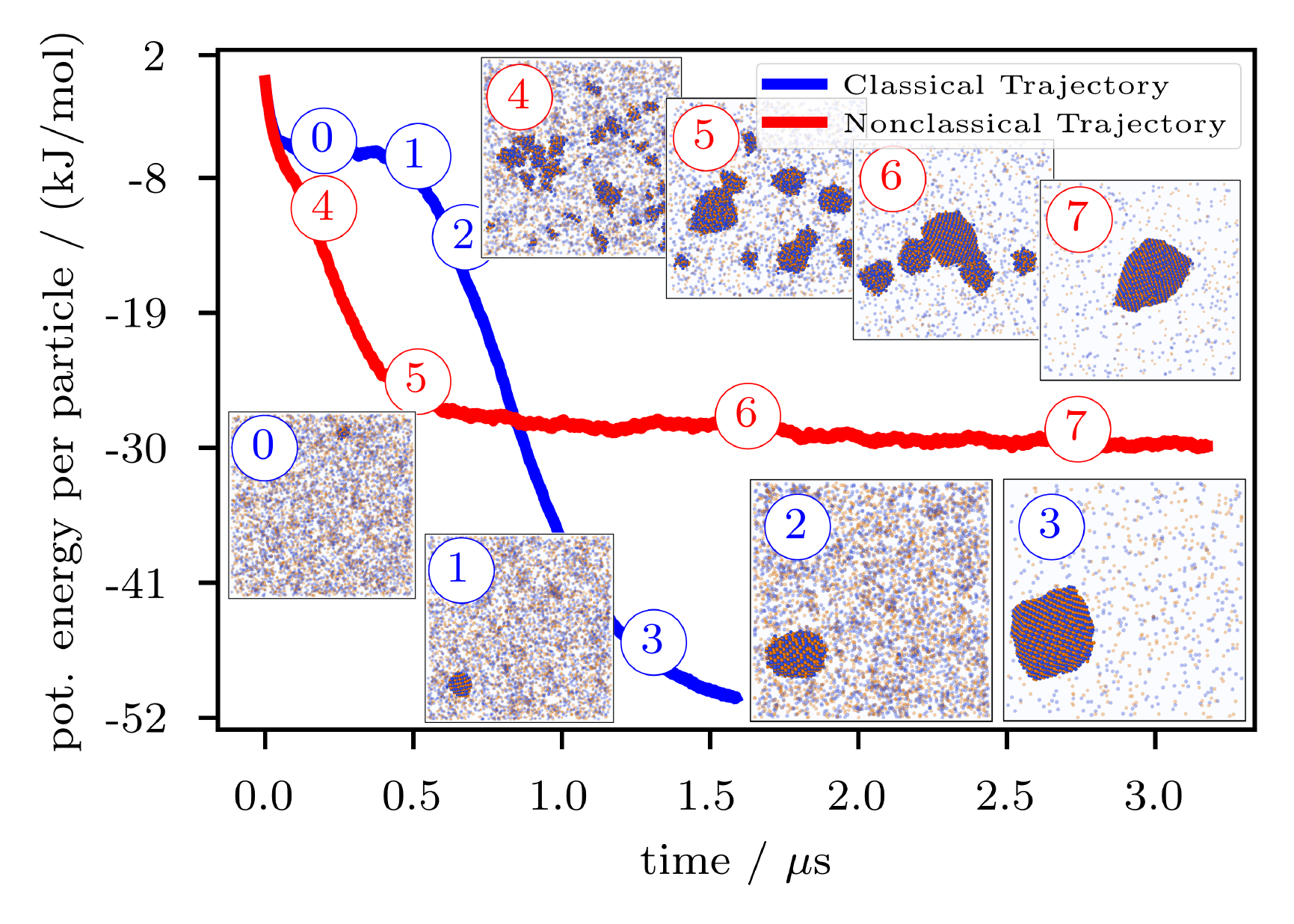}
\caption{Potential energy over the course of two PACSim simulations of a system of binary colloidal particles. Both simulations used $\num{6750}$ particles with a 1:1 ratio of positive and negative charges $\pm\qty{50.0}{\milli\volt}$. A classical, one-step nucleation trajectory (bottom, blue) is observed for positive particles of radius $r_P=\qty{97.5}{\nano\meter}$, negative particles of radius $r_N=\qty{105.0}{\nano\meter}$, and Debye length $\lambda_D=\qty{5.0483}{\nano\meter}$. A non-classical, two-step nucleation trajectory (top, red) is observed for $r_P=\qty{85.0}{\nano\meter}$, $r_N=\qty{105.0}{\nano\meter}$, and $\lambda_D=\qty{5.2}{\nano\meter}$.}
\label{fig:IV.A_nucleation}
\end{figure}

Slight alteration of simulation conditions for a system of binary colloidal particles interacting via the PACS potential can result in changes in the nucleation pathway that are consistent with experimental observations~\cite{Zang2025}. Accordingly, our simulations using PACSim show both classical nucleation---in which particles cluster into a crystalline seed that continues to grow---and non-classical nucleation---in which particles aggregate, and the aggregates then transition into crystalline structures (see \cref{fig:IV.A_nucleation} for snapshots from two different simulation trajectories undergoing classical or non-classical nucleation).

\subsection{Stability}
\label{sec:stability}

\begin{figure}[tb]
\centering
\includegraphics[width=\linewidth]{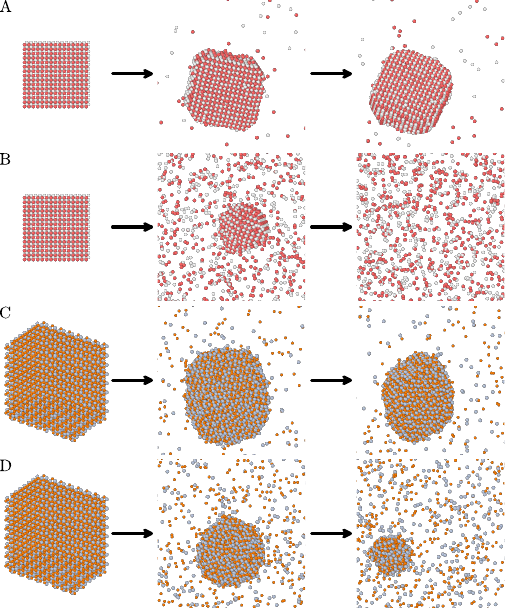}
\caption{Stability of crystalline seeds during $\sim\qty{157}{\nano\second}$ MD simulations at room temperature $T=\qty{298}{\kelvin}$ for different Debye lengths $\lambda_D$. Snapshots show the initial crystal structure (left), the configuration halfway through the simulation (middle), and the final configuration (right). (A)~Stable \ce{CsCl}-like crystal seed at $\lambda_D=\qty{5.0}{\nano\meter}$. (B)~Melting \ce{CsCl}-like crystal seed at $\lambda_D=\qty{4.8}{\nano\meter}$. (C)~Stable \ce{Th3P4}-like crystal seed at $\lambda_D=\qty{5.2}{\nano\meter}$. (D)~Slowly melting \ce{Th3P4}-like crystal seed at $\lambda_D=\qty{5.1}{\nano\meter}$.}
\label{fig:stability}
\end{figure}

The \texttt{pacsim-create} utility allows to initialize simulations from a specific crystal structure (see \cref{sec:pacsim-create}). As an example, we generate initial configurations from \ce{CsCl} and \ce{Th3P4} crystal structures provided in CIF format, which are rescaled to the colloidal length scale using the \texttt{LatticeBuilder} so that the sum of the polymer-brush repulsion $V_P$ and electrostatic interaction $V_E$ is minimized at a given Debye length $\lambda_D$ (see \cref{fig:stability}). In \texttt{pacsim-run}, we then monitor the stability of the crystalline seed in a longer simulation at room temperature $T=\qty{298}{\kelvin}$. 
As expected, the crystals remain stable at larger Debye lengths $\lambda_D$, whereas they melt at smaller $\lambda_D$. As discussed in Ref.~\cite{vankesteren-surfactant-2026}, this type of simulation probes the stability of a prescribed structure, but it does not by itself establish whether that structure is kinetically accessible during self-assembly.

\subsection{Reseeding}
\label{sec:Reseeding}

\begin{figure}[t]
\begin{yamlcodebox}{reseeding.yaml}
final_modifiers:
  - "SeedModifier"
  - "SeedModifier"

final_modifiers_parameters:
  - seed_filename: CsCl.gsd
    overlap_distance: !Quantity
      unit: nanometer
      value: 230.0
    seed_fractional_position: [|0.25|, |0.25|, |0.25|]
  - seed_filename: Th3P4.gsd
    overlap_distance: !Quantity
      unit: nanometer
      value: 230.0
    seed_fractional_position: [|0.75|, |0.75|, |0.75|]

...  # Remaining part of configuration file.
\end{yamlcodebox}
\caption{Configuration file excerpt for \texttt{pacsim-create} that places two different crystalline seeds from separate simulations into the initial configuration. 
The resulting seeded system and its subsequent MD evolution are shown in \cref{fig:seeding}.}
\label{fig:reseeding.yaml}
\end{figure}

\begin{figure}[t]
\centering
\includegraphics[width=\linewidth]{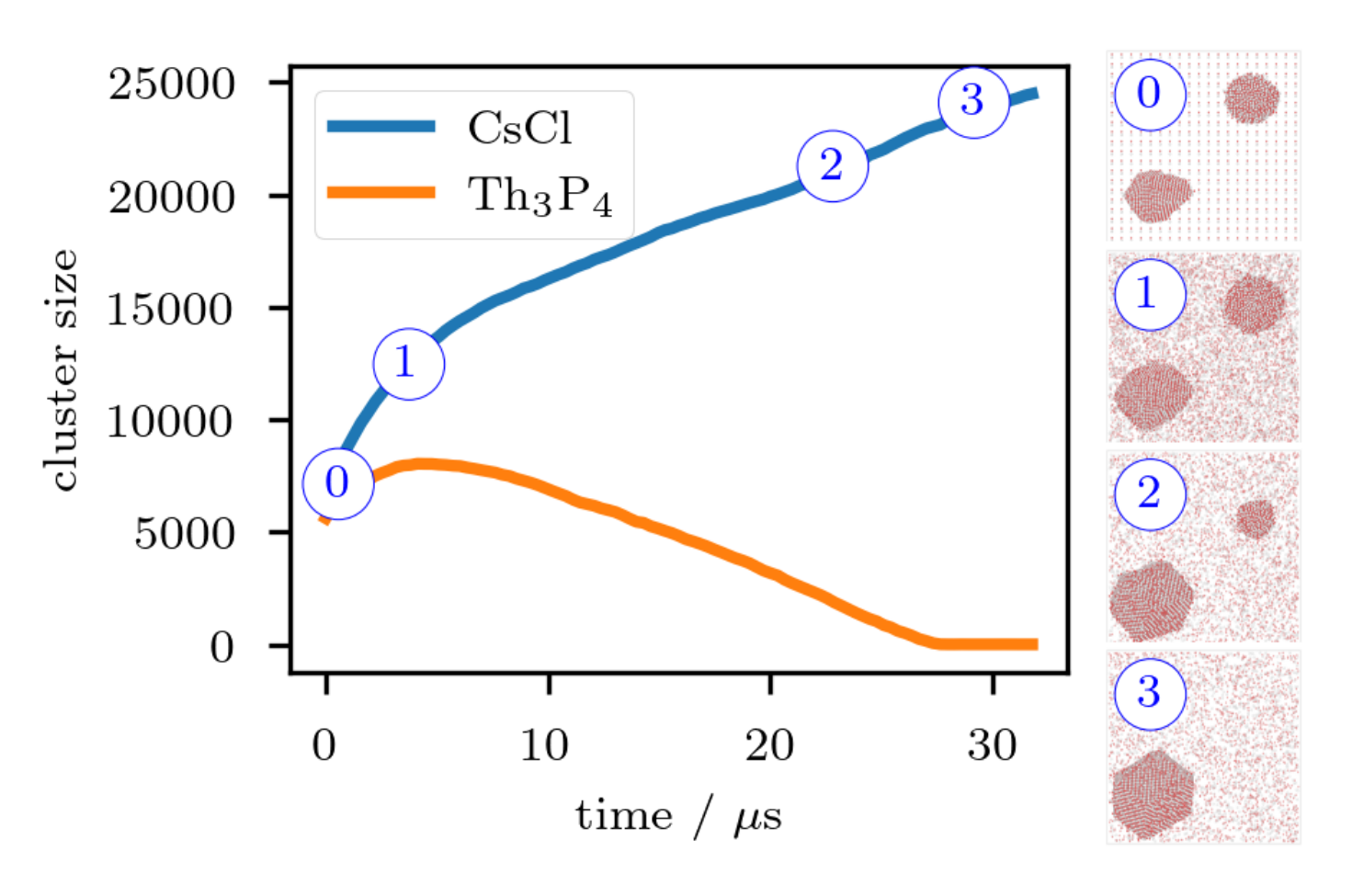}
\caption{Cluster sizes during an MD simulation seeded with both \ce{CsCl}-like and \ce{Th3P4}-like crystals 
at $\lambda_D=\qty{5.05}{\nano\meter}$. Numbered snapshots correspond to the initial configuration (generated with the configuration file of \cref{fig:reseeding.yaml}), the early growth of both crystals, the subsequent shrinkage of the \ce{Th3P4}-like crystal, and its eventual disappearance from the simulation. Note, this system is approximately 15-times more dense than in Fig.~\ref{fig:stability}D, hence both crystals can grow in this condition when not competing.}
\label{fig:seeding}
\end{figure}

Starting simulations from a crystalline seed provides a direct way to assess crystal stability by determining whether the seed grows or melts under a given set of conditions. Furthermore, seeded simulations can be used to compare the stability of competing crystal structures and identify which structure is more favorable for the chosen simulation parameters. To enable these studies, we implemented a seeding feature that allows an arbitrary set of particles to be overlaid when an initial configuration file is created.

Three parameters are necessary in the \texttt{pacsim-create} configuration file to seed a simulation: the GSD files that contain the seed configurations, the locations for each seed as the fractional coordinates of their centers, and an overlap tolerance for particles in the simulation (see \cref{fig:reseeding.yaml}). The seed configurations are merged with the base configuration generated by \texttt{pacsim-create} (for example, a gas of colloids created with the \texttt{ClusterGenerator}). Since extremely close particles can result in large forces and simulation instabilities, the overlap tolerance allows users to remove particles from the base configuration that have a surface-to-surface distance smaller than the overlap tolerance.

As an example, we embed two crystalline seeds with identical particle types into a gas of colloids (see \cref{fig:seeding}). The seeds were obtained from separate simulations where one formed a \ce{CsCl}-like crystal, and the other formed a \ce{Th3P4}-like crystal~\cite{Zang2025}. During the subsequent MD simulation, the two structures compete for the surrounding particles and, in this example, the CsCl-like crystal grows preferentially via Ostwald ripening until it alone survives.

\subsection{Explicit and Implicit Substrates}
\label{sec:Substrates}

\begin{figure}[t]
\begin{yamlcodebox}
    {explicit\_substrate.yaml (\texttt{pacsim-create})}
masses:
  "S": !Quantity
    unit: dalton
    value: 0.0
  ...  # Masses of non-substrate particles.
radii:
  "S": !Quantity
    unit: nanometer
    value: 30.0
  ...  # Radii of non-substrate particles.
surface_potentials:
  "S": !Quantity
    unit: millivolt
    value: -50.0
  ...  # Charges of non-substrate particles.
  
initial_modifiers: 
    - "SubstrateModifier"

initial_modifiers_parameters:
  - "substrate_type": "S"

...  # Remaining part of configuration file.
\end{yamlcodebox}

\begin{yamlcodebox}
    {implicit\_substrate.yaml (\texttt{pacsim-run})}
use_implicit_substrate: true
substrate_wall_charge: !Quantity
    unit: millivolt
    value: -50.0
...  # Remaining part of configuration file.
\end{yamlcodebox}
\caption{Configuration file excerpts for adding a substrate explicitly through a substrate modifier in \texttt{pacsim-create} (top) or implicitly through an additional interaction in \texttt{pacsim-run} (bottom).}
\label{fig:substrate.yaml}
\end{figure}

A charged substrate can promote heterogeneous nucleation and thereby guide crystallization of colloidal particles~\cite{Hueckel2020,Zang2025}. In PACSim, explicit substrates are generated by the substrate modifier of \texttt{pacsim-create}, which adds immobile substrate particles with zero mass and given radius and charge. In contrast, an implicit infinite-radius substrate particle is specified only in the configuration file for \texttt{pacsim-run} and reduces the computational cost relative to an explicitly modeled substrate layer at the expense of a less detailed representation of the substrate surface (see \cref{fig:substrate.yaml}).

\begin{figure}
\centering
\includegraphics[width=\linewidth]{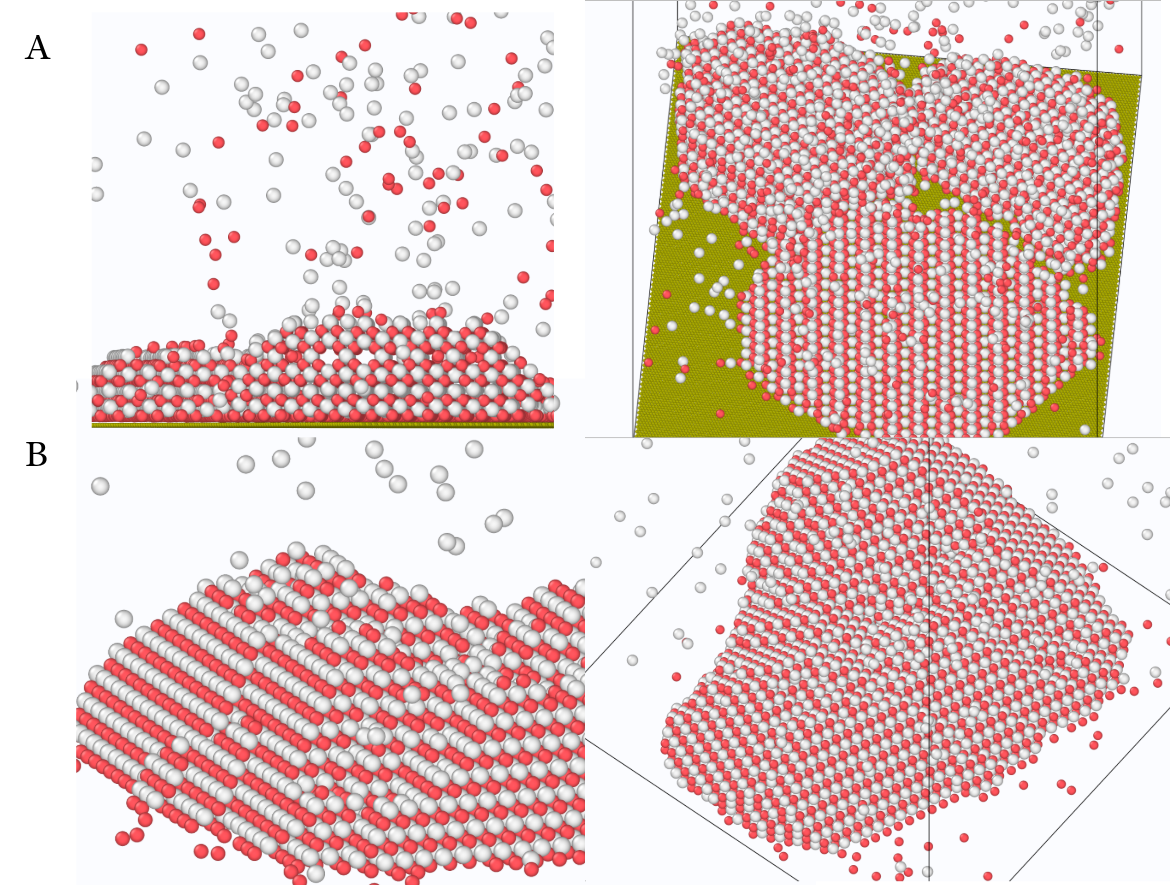}
\caption{Late-time snapshots from MD simulations with \texttt{pacsim-run} that include a negatively charged substrate at the bottom of the simulation box and gravity to bias particle motion toward the substrate surface. The two simulations of $\num{10000}$ mobile colloidal particles were performed under otherwise identical conditions and differ only in whether the substrate is modeled explicitly or implicitly. (A)~Explicit substrate representation with $\num{25864}$ immobile substrate particles of charge $\qty{-50}{\milli\volt}$, yielding an average GPU performance of $\sim\num{9200}$ MD steps per second. (B)~Implicit substrate representation with an infinite-radius substrate particle of charge $\qty{-50}{\milli\volt}$, yielding an average GPU performance of $\sim\num{16700}$ MD steps per second.}
\label{fig:substrates}
\end{figure}

We perform two MD simulations with \texttt{pacsim-run} under otherwise identical conditions, differing only in whether the substrate is modeled explicitly or implicitly (see \cref{fig:substrates}). Because the implicit substrate avoids the explicit simulation of substrate particles, it allows significantly more MD time steps per second. Since the substrate is represented differently in the two models, identical crystal structures are not expected. Indeed, crystals formed on the explicit substrate are flatter and contain both \ce{CsCl}-like and ``Zangenite’’ domains, consistent with an effectively higher surface charge as discussed in \citet{Zang2025}. 

\subsection{Update Reporters}

When the crystallization conditions are not known a priori, it can be useful to vary a control parameter during a simulation until crystallization is observed. In PACS, this can be used, for example, to mimic a dialysis protocol in which the salt concentration changes over time and thereby alters the Debye length $\lambda_D$. This functionality is offered in \texttt{pacsim-run} through update reporters (see \cref{fig:update.yaml}). By specifying a global parameter, an update function, and the duration over which the update is applied, one can prescribe time- and state-dependent changes to simulation parameters.

As an example, we compare two simulations with identical starting conditions in which the Debye length $\lambda_D$ is updated dynamically using different protocols (see \cref{fig:dialysis}). In the first case, $\lambda_D$ is increased according to a simple ramp. In the second case, the ramp is halted once a cluster containing $20$ colloidal particles is detected. The first protocol reaches larger values of $\lambda_D$, where strong interparticle attractions quickly result in the nucleation of all particles into small clusters. In contrast, the smaller final value of $\lambda_D$ in the second protocol suppresses additional nucleation events, allowing the crystals that have already nucleated to grow larger.

\begin{figure}[t]
\begin{yamlcodebox}{update\_reporter.yaml}
update_reporter: RampUpdateReporterUntilCluster

update_reporter_parameters:
  # Update Debye length from 4.5 nm to maximum 5.5 nm.
  parameter_name: debye_length
  start_value: !Quantity
    unit: nanometer
    value: 4.5
  end_value: !Quantity
    unit: nanometer
    value: 5.5
    
  # Update every 1000000 steps over 100000000 steps.
  final_update_step: 100000000
  update_interval: 1000000
  
  # Check for clusters every 1000000 steps.
  # Stop updating once 20 particles clustered.
  # Particles in cluster are less than 230 nm apart.
  check_interval: 1000000
  cluster_size: 20
  cutoff_distance: !Quantity
    unit: nanometer
    value: 230.0
  
  # Store time step and current value of parameter.
  filename: ramp_update_reporter.csv
  print_interval: 1000000

...  # Remaining part of configuration file.
\end{yamlcodebox}
\caption{Configuration file excerpt for \texttt{pacsim-run} that enables an update reporter which increases the Debye length $\lambda_D$ until a cluster containing at least $20$ colloidal particles is detected.}
\label{fig:update.yaml}
\end{figure}

\begin{figure}[t]
\centering
\includegraphics[width=\linewidth]{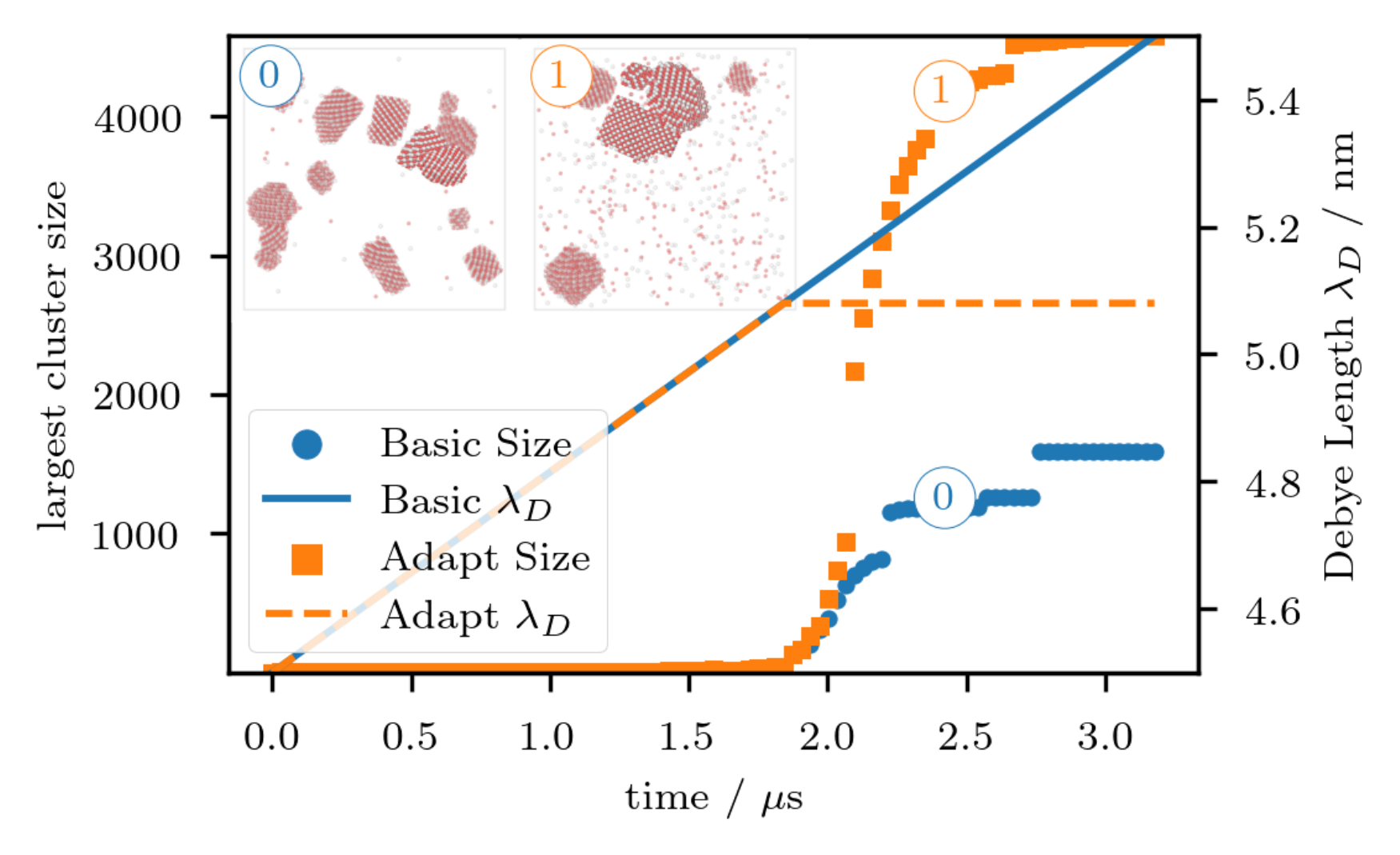}
\caption{Debye length $\lambda_D$ and largest cluster size over the course of two MD simulations with identical starting conditions. One simulation increases $\lambda_D$ over the entire simulation whereas the other one halts ramping $\lambda_D$ once a cluster containing at least $20$ colloidal particles is detected (see \cref{fig:update.yaml}).}
\label{fig:dialysis}
\end{figure}

\subsection{Clusters}

The \texttt{pacsim-create} utility supports the generation of initial configurations with constrained clusters of particles that evolve as rigid composite objects during the MD simulation with the \texttt{pacsim-run} utility. Such preassembled colloidal clusters are a common strategy to access specific crystal structures~\cite{Dijkstra2013,Ducrot2017,he2020colloidal}. 

\begin{figure}[t]
\centering
\includegraphics[width=\linewidth]{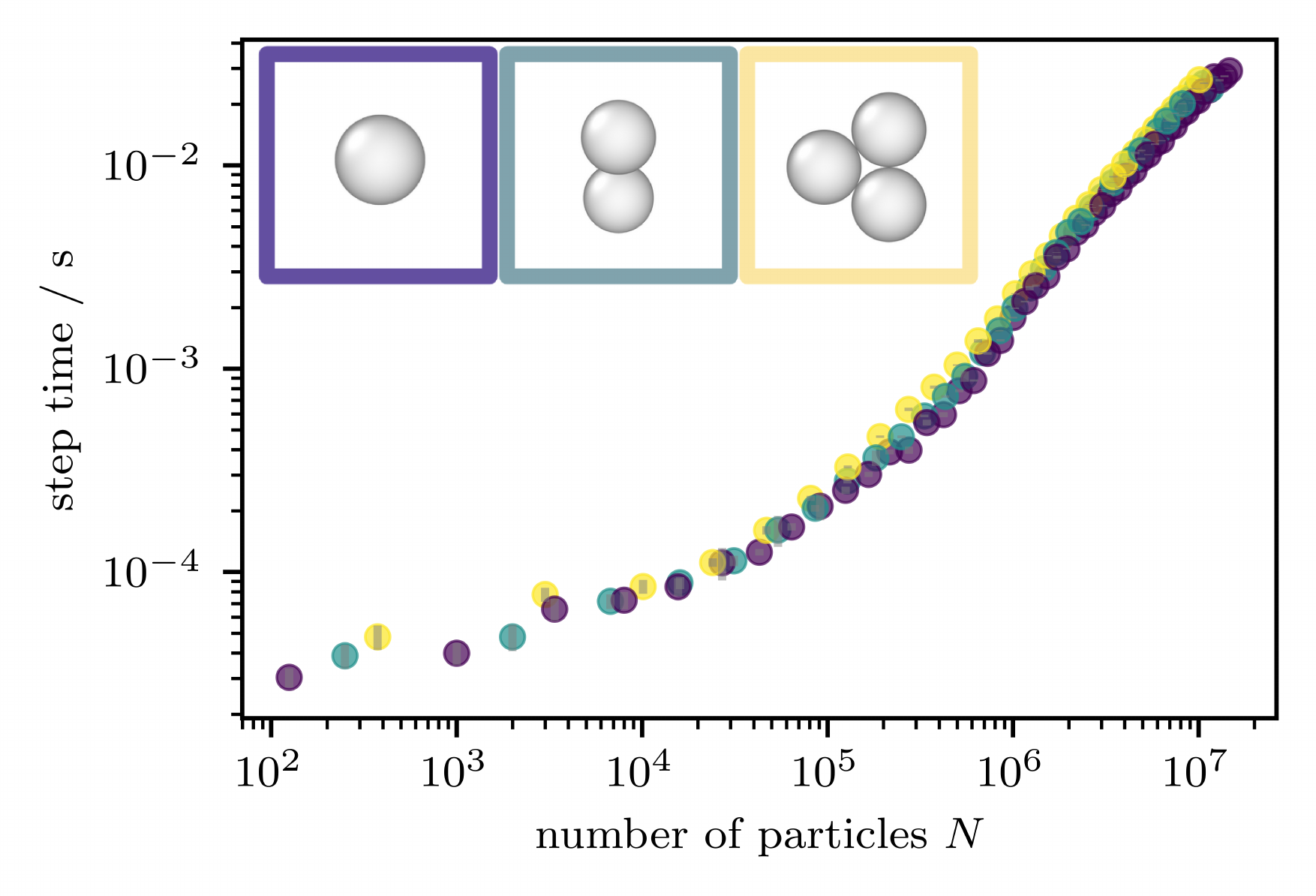}
\caption{MD step time as a function of number of particles for three different colloidal clusters as shown in the inset: a single particle, a two particle dumbbell, and three particles in an equilateral triangle.}
\label{fig:scaling}
\end{figure}

To demonstrate the use of this feature in PACSim, we consider simulations of dimers (dumbbells) and planar trimers (see inset of \cref{fig:scaling}). Simulations are performed in bulk at fixed molecular number density using many copies of each cluster type. Because distance constraints may have a significant impact on computational performance, these systems provide a useful benchmark for PACSim. For both individual particles and small constrained clusters, the simulation speed scales sublinearly until a system size of $\sim10^{5}$ particles, after which the scaling becomes approximately linear (see \cref{fig:scaling}). Introducing pairwise distance constraints reduces the sparsity of the constraint matrix, which increases the cost of the sparse matrix multiplications in the Constant Constraint Matrix Approximation algorithm of OpenMM~\cite{Eastman2010}. Consequently, clusters with more constraints exhibit a slightly larger prefactor in the overall scaling relation.

\subsection{Enhanced Sampling with PLUMED}

PLUMED is a plugin for MD simulations that allows for on-the-fly calculation of collective variables and enhanced-sampling simulations~\cite{plumed-consort-2019}. OpenMM interfaces with PLUMED through the \texttt{openmm-plumed} plugin. To enable PLUMED in PACSim, a PLUMED control script is specified through the \texttt{plumed\_script} keyword in the YAML configuration file for \texttt{pacsim-run} (see \cref{fig:run.yaml}).

\begin{figure}[t]
\centering
\includegraphics[width=\linewidth]{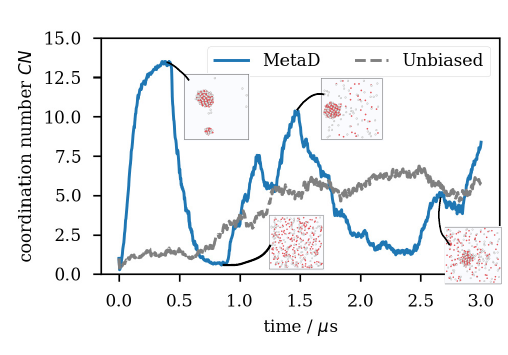}
\caption{Average coordination number during a well-tempered metadynamics simulation of a 432-particle binary colloidal system, using the mean particle coordination number as the biased collective variable. An unbiased simulation performed under otherwise identical conditions is shown for comparison. Representative snapshots illustrate the repeated crystallization and dissolution events observed during the biased simulation.}
\label{fig:metad}
\end{figure}

As an example, we use PLUMED to perform a metadynamics simulation~\cite{bussi2020using} under conditions where spontaneous nucleation is normally not observed. Several related examples of using PLUMED to assess crystal stability are available in the PLUMED tutorials~\cite{tribello2025tutorials}. Here, we bias the mean coordination number of all particles (see \cref{fig:metad}). We observe repeated crystallization and dissolution events over a $\qty{3}{\micro\second}$ simulation. In contrast, an unbiased simulation aggregates but does not crystallize over the same time interval. We note that using PLUMED incurs a large performance cost, hence we show results for a relatively small system size. Nevertheless, enhanced-sampling techniques produce exponential speedups for problems involving rare events, and the performance gain can easily outweigh the additional computational overhead. We also note that OpenMM has its own implementation of metadynamics, which is more performant but has a much more limited set of collective variables and other biasing techniques.

\section{Conclusions}
\label{sec:conclusions}
As illustrated by the examples in \cref{sec:cookbook}, PACSim provides a flexible platform for studying the crystallization of charged colloidal particles in aqueous solution. The underlying OpenMM engine is sufficiently fast and scalable to enable simulations of large systems and very slow crystallization processes on a single GPU. We have demonstrated several techniques that are highly valuable for exploring the wide range of possible structures that can be formed through the PACS approach. Incorporating enhanced-sampling approaches and complex crystal descriptors from PLUMED allows us to determine which crystal types are stable under particular conditions. An automated feedback loop between changing interaction parameters and current cluster size through PACSim's update reporters has proven to be a valuable tool for efficiently identifying conditions that promote large-crystal growth while limiting secondary nucleation. Finally, the ability to perform seeded growth further allows for tests of crystal stability and the formation of very large faceted crystals comparable to those observed experimentally.

To date, we have performed all of our simulations with a PACS potential described by \cref{eq:brush,eq:el}. This has been sufficient to get qualitative, and in some cases quantitative, agreement with experiments, in particular with regards to the type of structure and the crystallization pathway for a particular set of particle sizes and charges \cite{Hueckel2020,Zang2024,Zang2025,van2026light,vankesteren-surfactant-2026}. As described in \cref{sec:pacsim-run}, OpenMM makes it straightforward to implement additional analytical pair potentials. Future work could therefore explore whether more sophisticated expressions for the electrostatics [see \cref{eq:elect_full}] and polymer-brush repulsion improve agreement between simulated and experimentally observed crystallization conditions. At the same time, however, we must keep in mind that some model parameters, such as the brush density, are not known precisely, and changing those may have a comparable effect to changing the functional form of the potential. Careful experimental measurement of these quantities will therefore be important for properly assessing the quantitative accuracy of the PACS interaction model.

All of our studies to date have focused on the assembly from a dilute suspension, which is consistent with typical experimental protocols. Going forward, we would like to study bulk colloidal assembly. Since OpenMM is typically used for solvated biomolecular systems, it is well-suited for such studies and readily supports arbitrary triclinic periodic boxes and constant pressure simulations. To support these studies within PACSim, future versions would need to implement the ability to compress the box from a dilute to a dense system while equilibrating the starting configurations, as well as include initialization of periodic crystal structures whose simulation boxes match the corresponding unit cells.

In summary, PACSim simplifies the study of the self-assembly of charged colloidal particles in aqueous solution and, for our group, has substantially shortened the time between formulating new ideas and testing them \textit{in silico}. We hope that this open-source tool will prove useful to other researchers studying colloidal crystallization and related self-assembly problems, and we welcome future contributions of code, examples, and extensions.

\begin{acknowledgments}
We acknowledge Sanjib Paul who helped refine the simulation approach and contributed simulations to our previous publications. We thank NYU undergraduate Helen Zardus who helped implement gravity in PACSim. We would also like to thank other members of the Sacanna lab, especially Theodore Hueckel and Shihao Zang, for their insights which shaped our modeling approaches.

This research was supported by the US Army Research Office under award number W911NF-21-1-0011 to SS and GMH. The work of PH, NS, and MSC was supported by the Simons Center for Computational Physical Chemistry at NYU (SCCPC, Simons Foundation Grant MPS-T-MPS-00839534, MET). SvK was supported by the Swiss National Science Foundation under the Postdoc Mobility Grant number 217966. Computational work was supported in part through the NYU IT High Performance Computing resources, services, and staff expertise, and simulations were partially executed on resources purchased by the SCCPC.
\end{acknowledgments}

\bibliography{pacsim}

\end{document}